\documentstyle[prd,aps,preprint]{revtex}
\begin{document}
\draft
\title{From Random Matrix Theory to Statistical Mechanics - Anyon Gas}
\author{ Daniel ${\rm Alonso}^{1,2}$ and Sudhir R. ${\rm Jain}^{2,3}$ \\
(1) Department of Fundamental and Experimental Physics, \\
University of La Laguna, La Laguna 38204 \\
Tenerife, Canary Islands, Spain \\
(2) Facult\'e des Sciences and \\
Center for Nonlinear Phenomena and Complex Systems, \\
 Universit\'e Libre de Bruxelles,
Campus Plaine C.P. 231, \\ Boulevard du Triomphe, 
1050 Bruxelles, Belgium \\
(3) Theoretical Physics Division, Bhabha Atomic Research Centre,\\
 Bombay 400085, India}

\maketitle

\begin{abstract}

Motivated by numerical experiments and studies of quantum 
systems which are classically chaotic, we take a Random Matrix description of 
a Hard-sphere gas to Statistical Mechanical description. We apply this to Anyon gas 
and obtain a formal expression for the momentum distribution. Various limiting 
situations are discussed and are found in agreement with the well-known results on
Hard-sphere gas in low-density regime. 
\end{abstract}

\narrowtext

Random Matrix Theory and Statistical Mechanics are employed for the 
description of many-body systems like nuclei, metallic clusters and so on. During
the last decade or so, it has become increasingly clear that spectral fluctuations 
of simple quantum systems \cite{1} whose classical counterpart are chaotic, as well 
as the eigenfunctions, have features \cite{2} in striking agreement with random matrix 
theory (RMT). Originating from a conjecture of Berry \cite{3} regarding the eigenfunctions as a 
random superposition of plane waves, it was shown that a contact with statistical 
mechanics can be realised \cite{4}. In this Letter, we show that this connection is indeed 
a way to relate RMT and Statistical Mechanics.

It is well-known that statistical mechanics in two dimensions is at the core of 
several phenomena of great interest such as the fractional Hall effect, high-$T_c$ 
superconductivity, and others \cite{5,6}. Unlike in three dimensions the analysis 
here is plagued with difficulties stemming from an unusual, non-local exchange interaction 
potential. As a result identical particles in two dimensions exhibit 
statistics interpolating between the Fermi-Dirac and Bose-Einstein distributions, 
generically these are termed as Anyons \cite{7}. In this Letter, we address this 
fundamental problem in two dimensions and cast it systematically employing an ansatz 
for the eigenfunctions consistent with the connection mentioned above, and incorporate
the Braid group governing exchange symmetries. We show a way to arrive at the momentum 
distribution upto $O(\hbar^2)$ and express it in terms of 
quantities involving the counting problem in Braid group. Statistical mechanics of
Anyons has been studied in the past \cite{8} and it is very interesting to see that
the second virial coefficient obtained then and now, 
give rise to a Sum Rule. However, in marked distinction with previous works, 
calculation of higher-order virial coefficients
is also possible here - a facet which makes the present approach a novel alternative. To this end, we argue that the third virial coefficient 
 has the 
form in agreement with some recent results. We re-iterate that the aim of this 
Letter is to arrive at quantum statistical mechanics without having to make a hypothesis about thermal bath, and that the present approach is guided by random matrix theory.

The fundamental hypotheses of RMT are (a) the hamiltonian of a 
given system belongs to an ensemble of hamiltonians, and (b) real quantum 
mechanics is enough for description of physical systems if time-reversal 
symmetry is taken into account appropriately \cite{9}. It is the latter of
the two that leads to exactly three universality classes in the 
RMT - Orthogonal (even spin, time-reversal invariant, OE) Unitary 
(time-reversal non-invariant, UE) and Symplectic Ensembles (odd spin, 
Time-Reversal Invariant, SE) \cite{10}. Owing to the spin-statistics connection 
in two dimensions \cite{11}, we know that there is fractional statistics 
and fractional spin. In the context of symmetries, in two dimensions, parity (P) 
and time-reversal (T) both are broken.

Let us consider a system of $N$ hard spheres ('discs' in two dimensions), 
each of radius $a$, enclosed in a box of edge-length $L+2a$. Centres of two hard 
spheres $\vec x_i$ and $\vec x_j$ are such that $|\vec x_i-\vec x_j |\ge 2 a$. 
The canonical pair of coordinates describing these particles are $(\vec X, \vec P)$ where 
$\vec X= (\vec x_1,\vec x_2, \cdots, \vec x_N), \, 
\vec P =(\vec p_1,\vec p_2, \cdots, \vec p_N)$. Energy eigenfunctions, 
$\psi_{\alpha}(\vec X)$ corresponding to eigenvalue $E_{\alpha}$ vanish on the 
boundary of the enclosure. A typical eigenfunction is irregular, with a Gaussian amplitude 
distribution and the spatial correlation function of the same is consistent with the 
conjecture of Berry which allows us to represent this eigenfunction as a superposition 
 \cite{4} :
\begin{equation}
\label{F1}
\psi_{\alpha}(\vec X)= N_{\alpha} \int d^{dN}\vec P A_{\alpha}(\vec P) 
\delta (P^2-2mE_{\alpha})\, e^{\frac{i}{\hbar}\vec X \cdot \vec P}
\end{equation}
\noindent
with $N_{\alpha}$ given by the normalization constant, and $A_{\alpha}'s$ satisfying 
the two-point correlation function
\begin{equation}
\label{F2}
\big< A_{\alpha}^*(\vec P) A_{\gamma}(\vec P') \big>_{ME} = \delta_{\alpha \gamma}
\frac{\delta^{dN}(\vec P - \vec P')}{\delta (\vec P^2-\vec P'^2)}, 
\end{equation}
\noindent
$d$ denotes the number of coordinate-space dimensions. The average in (\ref{F2}) 
is a matrix-ensemble (ME) average which originates from the fact that the hamiltonian, 
$H$ of the system belongs to an ensemble of matrices satisfying associative  division 
algebra \cite{10,frob} in consistency with quantum mechanics. The eigenstate ensemble (EE) used in \cite{4} 
is nothing but a consequence of underlying matrix ensemble in RMT, the eigenfunctions 
then satisfy all the properties numerically observed and analytically represented 
in (\ref{F1}), (\ref{F2}) \cite{12}. The correlation functions (\ref{F2}) decide whether 
time-reversal symmetry is preserved $(A_{\alpha}^*(\vec P)=A_{\alpha}(-\vec P))$ or 
broken $(A_{\alpha}^*(\vec P) \neq A_{\alpha}(-\vec P))$, accordingly the corresponding 
matrix ensemble belongs to OE or UE respectively. As noted in \cite{4}, the higher-order 
even-point correlation functions factorize and the odd-ones vanish. A very important aspect of the ansatz 
(\ref{F1}), (\ref{F2}) is that the Wigner function corresponding to $\psi_{\alpha}(\vec X)$ is microcanonical, 
or, is proportional to $\delta (H-E_{\alpha})$ which, in a sense, incorporates ergodicity. 
We note here that, starting from an ansatz very similar to the one above, it is possible 
to obtain the quantum transport equation \cite{13} where it is important to 
relate a given quantum state with the admissible energy surface in phase 
space; thus the above ansatz is in conceptual agreement with the ergodic aspect of many-body system. 
Moreover, this choice fixes the Thomas-Fermi density of states naturally. It now becomes 
important to emphasize that we must restrict ourselves to dilute gas of hard-spheres 
and also assume that the size of sphere is much lesser than the thermal de Broglie 
wavelength. Thus, the ansatz establishes, in fact, a link between RMT and statistical mechanics. We now incorporate 
the case of two dimensions which otherwise presents enormous difficulties.

In two dimensions, the solutions of the Schr\"odinger equation, 
$\psi(\vec x_1,\vec x_2, \cdots, \vec x_N)$, under an exchange of two coordinates of particles 
satisfies
\begin{eqnarray}
\label{F3}
&& \psi(\vec x_1,\cdots, \vec x_i, \cdots, \vec x_j,\cdots , \vec x_N) \nonumber \\
= e^{i \pi \nu}
&& \psi(\vec x_1,\cdots, \vec x_j, \cdots, \vec x_i,\cdots , \vec x_N)
\end{eqnarray}
\noindent
where $\nu$ is arbitrary and defines statistics. For $\nu=0$ and $\nu=1$, with 
(\ref{F2}), one gets the Bose-Einstein and Fermi-Dirac distributions. This non-trivial phase 
and the resulting boundary condition arises from the fact that the effective 
configuration space, $M_{N}^2$ has a fundamental group, $\pi_1(M_{N}^2)=
B_N$ \cite{14}, the Braid group of $N$ objects which is an infinite, non-abelian 
group. $B_N$ is generated by $(N-1)$ elementary moves $\sigma_1, \cdots, \sigma_{N-1}$ 
satisfying the Artin relations,
\begin{eqnarray}
\label{F4}
\sigma_i\sigma_{i+1}\sigma_i=\sigma_{i+1}\sigma_i\sigma_{i+1} \, \, 
(i=1,2,\cdots, N-2) \nonumber \\
\sigma_j\sigma_i=\sigma_i\sigma_i, \, \, |i-j| \ge 2 
\end{eqnarray}
\noindent
the inverse of $\sigma_i$ is $\sigma_i^{-1}$, the identity is denoted by 
${\sl I}$, and the centre of $B_n$ is generated by $(\sigma_1\sigma_2 \cdots 
\sigma_{N-1})^N$. The multivaluedness of the eigenfunction originates from the phase 
change in  effecting an interchange between two coordinates $x_i^{(1)}$ and 
$x_i^{(2)}$ (superscripts refering to components) which can be expressed as
\begin{eqnarray}
\label{F5}
V=\exp(i \nu \sum_{i<j} \phi_{ij}), \nonumber  \\
\phi_{ij}=\tan^{-1}\big(\frac{x_i^{(2)}-x_j^{(2)}}{x_i^{(1)}-x_j^{(1)}} \big).
\end{eqnarray}

The description adopted by us here is referred to as the Anyon Gauge. It is important 
to realise that a set of coordinate configuration can be reached starting from some initial 
coordinates of $N$ particles in an infinite ways, each possibility manifested by an
action of an element $\beta \in B_N$.

The connection between initial and final sequences is given by (\ref{F3}), via 
the character $\chi(\beta)$ of the specific element. Thus, to every $\beta \in 
B_N$, we can associate the affected partial amplitude
 $\psi_{\alpha}(\beta:\vec x)$ \cite{15}. With one-dimensional unitary representation of the braid group, the 
rudiments of quantum mechanics allow us to write
\begin{equation}
\label{F6}
\Phi_{\alpha}(\vec X)=\sum_{\beta \in B_n} \chi(\beta) 
\psi_{\alpha}(\beta :\vec X)
\end{equation}
\noindent
where $\psi_{\alpha}(\beta :\vec X)$ is the probability amplitude associated in changing a 
configuration $\vec X$ to $(\beta :\vec X)$ - a configuartion after the action of $\beta$ on $\vec X$. 
The wavefunction $\Phi_{\alpha}(\vec X)$ is to be understood as appropriately 
normalised. The ansatz for $V \psi_{\alpha}(\beta :\vec X)$ is now
\begin{equation}
\label{F7}
V \psi_{\alpha}(\beta :\vec X)=
N_{\alpha} \int d^{2N}\vec P A_{\alpha}(\beta: \vec P) 
\delta (P^2-2mE_{\alpha}) e^{\frac{i}{\hbar}\vec X \cdot \vec P}
\end{equation}
\noindent
with $A_{\alpha}(\beta: \vec P)$ satisfying
\begin{equation}
\label{F8}
\big< A_{\alpha}^*(\beta_1:\vec P_1) A_{\gamma}(\beta_2:\vec P_2) \big>_{ME} =
 \delta_{\alpha \gamma}
\frac{\delta^{2N}\big((\beta_1:\vec P_1) - (\beta_2:\vec P_2)\big)}
{\delta (\vec P^2_1-\vec P^2_2)},
\end{equation}
\noindent
$(\beta_1,\beta_2 \in B_N)$, and $A_{\alpha}(\vec P)$ satisfy the twisted boundary
conditions,
\begin{eqnarray}
\label{F9}
&& A_{\alpha}(\vec p_1,\cdots, \vec p_i, \cdots, \vec p_j,\cdots , \vec p_N) \nonumber \\ 
 = e^{i \pi \nu}
&& A_{\alpha}(\vec p_1,\cdots, \vec p_j, \cdots, \vec p_i,\cdots , \vec p_N)
\end{eqnarray}

The question now is in specifying exactly what the 
matrix ensemble is in this case? The form of (\ref{F8}) with $A_{\alpha}$'s not 
restricted to real, takes into account the T-breaking, and (\ref{F9}) makes the 
ensemble handed or chiral as a result of P-breaking. Thus (\ref{F7})-(\ref{F9}) gives 
the complete description and the ME is, in fact, the chiral-Gaussian Unitary Ensemble 
(ch-GUE) \cite{16}. It can be easily shown that the Wigner distribution is
\begin{eqnarray}
\label{10}
\big<\rho_{\alpha}^W(\vec X , \vec P) \big>_{ME}=n_{\alpha}^{-1} h^{-2N} 
\delta(\frac{P^2}{2m}-E_{\alpha}), \nonumber \\
n_{\alpha}= \frac{1}{N! \Gamma(N) E_{\alpha}}\big( 
\frac{m L^2 E_{\alpha}}{2 \pi \hbar^2} \big)^N
\end{eqnarray}

For the momentum distribution, we need to evaluate the ME-average of 
$\tilde \Phi_{\alpha}^*(\vec P)\tilde \Phi_{\gamma}^*(\vec P")$ with $\tilde \Phi 
\equiv V\Phi_{\alpha}$. With the above ansatz and conditions supplementing it, this average is
\begin{eqnarray}
\label{F11}
&& {\cal{F}}(\vec P)= 
\big< \tilde \Phi_{\alpha'}^*(\vec P)\tilde \Phi_{\gamma}^*(\vec P") \big>_{ME} = \nonumber \\
&& h^{2N}\delta_{\alpha' \gamma} N_{\alpha '} N_{\gamma} 
\sum_{n,m=0} \sum_{\beta_1(m)} \sum_{\beta_2(n)} \chi^*(\beta_1) \chi(\beta_2) 
\delta (P^2-2mE_{\alpha '}) \nonumber \\ 
&& \times \delta_{\cal D}^{2N} \bigg( \prod_{\alpha =0}^m 
\sigma_{\beta_1(\alpha )}^{\epsilon_{\beta_1}} \vec P"  -
      \prod_{\alpha =0}^n
\sigma_{\beta_2(\alpha )}^{\epsilon_{\beta_2}} \vec P 
\bigg) \bigg|_{\vec P = \vec P"} 
\end{eqnarray}
\noindent
where
\begin{equation}
\label{F12}
\delta_{{\cal D}}^{2N}(\vec Q)= h^{-2N} 
\int_{Domain, \cal D} d^{2N}X \exp \big( \frac{i}{\hbar} \vec Q \cdot \vec X \big) ;
\end{equation}
\noindent
$\vec P$ is identified with $\vec P"$ after the sum is performed.

With (\ref{F11}), the momentum distribution is given by
\begin{equation}
F(\vec p_1)= \frac{\int d \vec p_2 \cdots d \vec p_N {\cal F}(\vec P)}
                  {\int d \vec p_1 \cdots d \vec p_N {\cal F}(\vec P)}
\end{equation}
\noindent
which formally completes the deduction. However, an exact evaluation of this is 
very difficult and the difficulty is coming from  counting of irreducible 
words formed by the $\sigma 's$. To make the precise connection, we give results 
upto $O(\hbar^2/L^2)$, an order that is enough for second virial coefficient. Leaving 
the tedium of details to a later publication, we just give our 
result incorporating all elements leading to $O(\hbar^2/L^2)$,
\begin{eqnarray}
\label{F13}
&& F_2(\vec p_1)=(2 \pi m k T)^{-1} \exp \big( 
-\frac{\vec p_1^2}{2mkT} \big) \bigg\{ 1 \nonumber \\
&& + \big(\frac{h}{L}\big)^2 \frac{1}{2\pi m k T} 
\big( 2 e^{-\frac{\vec p_1^2}{2 m k T}} - 1 \big) G(N, \nu) +
O(\frac{h^4}{L^4}) \bigg\}
\end{eqnarray}
\noindent
where
\begin{equation}
\label{F14}
G(N, \nu) = \frac{\sum_{m=0}^{\infty} \sum_{K=-2m-1 (even)}^{2m+1} 
Q_K^{(m)}(N) \cos(\pi K \nu)   }
{1 + 2 \sum_{m=1}^{\infty} \sum_{K=-2m (odd)}^{2m}
P_K^{(m)}(N) \cos(\pi K \nu)   },
\end{equation}
\noindent
$Q_K^{(m)}$ is the number of elements in $B_N$ composed of $'m'$ generators whereby 
the momentum $\vec p_1$ is interchanged with another momentum yielding a character 
$\exp(i \pi K \nu)$ (or $\exp(-i \pi K \nu)$ since 
$Q_K^{(m)}(N)=Q_{-K}^{(m)}(N)$); $P_K^{(m)}(N)$ is the number of elements in $B_N$ 
contributing to identity with a character $\exp(i \pi K \nu)$ 
( or $\exp(-i \pi K \nu)$). Temperature is introduced above via the ideal gas law,
$E_{\alpha} = NkT_{\alpha}$.
 Unfortunately though, this counting problem stands open today
\cite{17}.  
It is very important to note that the ansatz (\ref{F7})-(\ref{F9}) for the special case when 
$\sigma^2_i=1$ for all $i$ where $B_N$ reduces to symmetric group, $S_N$, the well-known 
Fermi-Dirac and Bose-Einstein distributions follow. On evaluating pressure, $\Pi $ 
from (14), denoting area of the enclosure by A, we get 
${\Pi}A/kT = 1 - (2A)^{-1}{\lambda}^2G(N,{\nu})$, with ${\lambda}^2=h^2(2{\pi}mkT)^{-1}$.
We immediately see that $G(N,0)/(2N)$ and $G(N,1)/(2N)$ are $2^{-3/2}$ and $-2^{-3/2}$
respectively yielding the second virial coeffiecient for the Bose and Fermi gases 
\cite{18}. For the fractional case, with $\nu =$ even number,2j + $\delta $ ("boson-based
anyons"), comparing our result with \cite{8}, we get the Sum Rule mentioned in the
introduction:
\begin{equation}
-2^{-3/2}N^{-1}G(N,\nu ){\lambda}^2 = (-1+4|{\delta}|-2{\delta}^2){\lambda}^2/4,
\end{equation}
the right hand side belongs to \cite{8}.
It is important to note that our deduction is non-perturbative and in principle,
 we can get expressions for higher-order virial coefficients also \cite{19}.
To understand this, we observe that the relation (8) connects two momentum configurations of N particles, and not just the momenta of two particles. Thus, it contains information that can lead to all virial coefficients. For example, for the third virial coefficient, we need to evaluate contributions to $F(\vec p_1)$ when three momenta out of N are interchanged. The denominator of (13) contains those interchanges which braid three strands in such a way that the initial configuration of momenta is preserved whereas the numerator of (13) contains those which exchange the momentum assignment on all three strands. We have done the calculation and the third virial coefficient is expressible in terms of the specific counting problem of $B_N$. Here, in order to convince the reader, it suffices to make a comparative discussion with the existing calculation. For this, we write down the total contribution to the momentum distribution due to a triple interchange emerging from the elements of $B_N$ formed by M generators,
\begin{eqnarray}
\bigg({L \over \hbar}\bigg)^{2(N-2)}& &\sum_{-2M}^{2M}{2 \over 3} I_{2(N-3)}(2mE_{\alpha}-\vec p_1^2)\cos (\pi k\nu ){\cal R}_k^{M}(N) \nonumber \\ &+&2I_{2(N-3)}(2mE_{\alpha}-3\vec p_1^2)\cos (\pi k\nu ){\cal S}_k^{M}(N),
\end{eqnarray}
where ${\cal S}_k^{M}(N)$ (${\cal R}_k^{M}(N)$) are the number of elements of $B_N$ that (do not) change the momentum $\vec p_1$. $I_D(x)$ denotes the volume of a D-dimensional hyper-sphere of radius x. The reason we give this result here is to show that (17) is a Fourier series with harmonic terms like $\cos 2\pi \nu , \cos 4\pi \nu, $ and so on, in complete agreement with the conjectured form \cite{jm-ko}. It is becoming evident from the Monte Carlo calculations \cite{mash} that the third virial coefficient is a series with terms as $\sin ^2\pi \nu , \sin ^4\pi \nu $, and so on. Our formal result is thus in consonance with these works.     
Also, we mention that (16), (17) and the Monte Carlo estimates provide a non-trivial hint on the counting problem
itself.

To summarise, we have formulated in this Letter a way to approach Statistical Mechanics 
from RMT through the ansatz for eigenfunctions which is an essential 
dynamical input. Moreover, we have presented a first-principles-evaluation of 
the momentum distribution of a 'Hard-Sphere Anyon Gas' in the low-density regime 
which is of a great interest in current literature. The nature of difficulties, albeit 
well-known, are made explicit here in direct connection with the existing wisdom 
(or ignorance) of theory of braid groups. The parameter, $\nu $ has an analogous 
partner in quantum chromodynamics \cite{8} and we conjecture that the anyon gas 
discussed here and the $\nu \pi $-parametrised quantum chromodynamics belong to 
the same universality class of chiral-Gaussian Unitary Ensemble of RMT.

\section{Acknowledgements}
Authors express their deep debt of gratitude to Pierre Gaspard for
stimulating discussions and constructive criticism. S.R.J. is financially
supported by the "Communaute Francaise de Belgique" under contract no.
ARC-93/98-166.
\label{sectacknow}

\end{document}